\documentclass[%
 reprint,
 amsmath,amssymb,
 aps,
pra,
]{revtex4-1}

\usepackage{graphicx}
\usepackage{dcolumn}
\usepackage{bm}
\usepackage[mathlines]{lineno}

\begin{document}


\title{Establishing non-thermal regimes in pump-probe electron-relaxation dynamics}

\author{M. X. Na$^{1,2}$, F. Boschini$^{1,2}$, A. K. Mills$^{1,2}$, M. Michiardi$^{1,2,3}$,\\ R. P. Day$^{1,2}$, B. Zwartsenberg$^{1,2}$, G. Levy$^{1,2}$, S. Zhdanovich$^{1,2}$, \\A. F. Kemper$^{4}$, D. J. Jones$^{1,2\dagger}$, A. Damascelli$^{1,2\dagger}$\\
\normalsize{$^{1}$Quantum Matter Institute, University of British Columbia, Vancouver, BC, Canada V6T 1Z4}\\
\normalsize{$^{2}$Department of Physics and Astronomy, University of British Columbia, Vancouver, BC, Canada V6T 1Z1}\\
\normalsize{$^{3}$Max Planck Institute for Chemical Physics of Solids, 01187 Dresden, Germany}\\
\normalsize{$^{4}$Department of Physics, North Carolina State University, Raleigh, NC, 27695, USA}\\
\normalsize{$^\dagger$To whom correspondence should be addressed;}\\
\normalsize{E-mail:  djjones@physics.ubc.ca; damascelli@physics.ubc.ca}
}

\begin{abstract}
Time- and angle-resolved photoemission spectroscopy (TR-ARPES) accesses the electronic structure of solids under optical excitation, and is a powerful technique for studying the coupling between electrons and collective modes. One approach to infer electron-boson coupling is through the relaxation dynamics of optically-excited electrons, and the characteristic timescales of energy redistribution. A common description of electron relaxation dynamics is through the effective electronic temperature. Such a description requires that thermodynamic quantities are well-defined, an assumption that is generally violated at early delays. Additionally, precise estimation of the non-thermal window -- within which effective temperature models may not be applied -- is challenging. We perform TR-ARPES on graphite and show that Boltzmann rate equations can be used to calculate the time-dependent electronic occupation function $f(\epsilon, t)$, and reproduce experimental features given by non-thermal electron occupation. Using this model, we define a quantitative measure of non-thermal electron occupation and use it to define distinct phases of electron relaxation in the fluence-delay phase space. More generally, this approach can be used to inform the non-thermal-to-thermal crossover in pump-probe experiments.
\end{abstract}
\maketitle


\section{Introduction}
In recent years, the development of ultrashort laser pulses has enabled the study of many-body electron interactions and their intrinsic timescales in non-equilibrium conditions \cite{Dalconte2015, Giannetti2016}. Upon optical excitation, elementary scattering processes (electron-electron, electron-phonon, etc.) redistribute the laser energy absorbed, leading to a plethora of non-equilibrium phenomena, such as the melting of equilibrium phases \cite{Rohwer2011, Boschini2018}, the formation of metastable non-equilibrium phases \cite{Cocker2012, Wang2013}, and transient topological phases \cite{McIver2020}. The energy redistribution process is modulated by coupling strengths between all possible degrees of freedom, but predominantly by the electron-boson coupling, such as coupling to phonons and/or magnons \cite{Carpene2008, Tengdin2018}.

Commonly, the study of electron-boson coupling in pump-probe experiments invokes a two-temperature model (TTM), in that both electronic and bosonic populations may be described by their own, distinct, temperatures \cite{Kaganov1957, Anisimov1975, Maldonado2017}. In pump-probe thermomodulation experiments, the ultrashort, intense laser pulses create a non-equilibrium condition between the electrons and the lattice \cite{Fujimoto1984, Schoenlein1987, Elsayed-Ali1987}. Within the TTM description, the electron bath is assumed to be thermalized on the femtosecond time scale of the excitation: subsequent electron-boson scattering mediates the transfer of energy from the electron to the boson bath. The latter is also assumed to maintain a thermalized Bose-Einstein distribution via boson-boson scattering. The rate of energy transfer between electrons and bosons, and correspondingly the electrons' intrinsic relaxation times, are determined by the electron-boson coupling strength \cite{Anisimov1975}. This model was used in early pump-probe studies to extract the electron-phonon coupling in metals and BCS superconductors \cite{Allen1987, Brorson1990}. However, later experiments have shown that the electronic bath does not reach thermal equilibrium before electron-phonon scattering becomes relevant -- especially in the low excitation regime \cite{Groeneveld1992} -- thereby invalidating one of the key assumptions of the TTM \cite{Groeneveld1995, Rethfeld2002, Mueller2013}. Despite this, the TTM remains prominent in the analysis of pump-probe experiments \cite{Perfetti2007, Matsuzaki2009, Johannsen2013, Sobota2014a, Sterzi2016}. 

In systems where electrons couple to more than one bosonic mode, electron relaxation dynamics have been treated by using multi-temperature models (MTM), in which a distribution at finite temperature is used to describe each degree of freedom at every delay \cite{Bigot1996, Carpene2008, Patz2014}. Quasi-thermalized distributions of electrons, phonons, and magnons, have been successfully used to describe ultrafast demagnetization \cite{Bigot1996, Koopmans2010}, nematic fluctuations \cite{Patz2014}, orbital order \cite{Matsuzaki2009}, and electron-phonon coupling \cite{Allen1987}. In some cases, MTMs have been used to partition non-thermal distributions into independently thermalized sub-distributions. For example, the non-thermal phonon bath has been partitioned into strongly-coupled optical phonons (SCOPs) and the weakly-coupled lattice, which heats via anharmonic decay of the SCOPs \cite{Perfetti2007, Rettig2013}. The non-thermal electron bath has also been partitioned into two distributions with different chemical potentials and temperatures \cite{Gilbertson2012, Gierz2013c}, which are then fit with Fermi-Dirac (FD) distributions to extract the effective temperature for each sub-distribution.

Much effort has been made to describe the photo-excited electron distribution \cite{Carpene2006, Waldecker2016, Ono2018, Kemper2018}. While is it important to determine when temperature becomes a good description of the electronic distribution, electrons in the non-thermal regime are also rich with information, which we can retrieve via carefully designed pump-probe experiments. For instance, we have recently demonstrated how non-thermal features can be used to extract the mode-projected electron-phonon matrix element in graphite \cite{Na2019}. Here we take a closer look at the evolution of the whole electronic distribution. In TR-ARPES, the effort to distinguish between thermal and non-thermal electronic distributions is complicated by the following factors: (i) ARPES intensity is given by the spectral function, the photoemission matrix element, and the occupation function; the first two terms complicate the estimation of the effective electronic temperature via Fermi-edge fitting \cite{Stange2015}. (ii) The definition of the (multiple) phonon temperatures are arbitrary, as TR-ARPES does not access phonon occupation directly. 

In this work, we circumvent the concept of temperature and explore the evolution of the electronic and bosonic populations within the framework of Boltzmann rate equations. This well-established methodology has been successful in describing electron dynamics in metals \cite{Rethfeld2002, Kabanov2008, Mueller2013}, reproducing dynamical trends in time-resolved reflectivity \cite{Groeneveld1995} and electron diffraction \cite{Waldecker2016}. We demonstrate that the consideration of electron-electron (e-e), electron-phonon (e-ph), and phonon-phonon (ph-ph) scattering can qualitatively reproduce key non-thermal features in the electron distribution, as measured by TR-ARPES on graphite. Finally, we simulate the evolution of the electron distribution as a function of pump fluence. By defining a quantitative measure of non-thermal electron occupation, we identify distinct phases in the fluence-delay phase space in which the electron distribution either does or does not manifest non-thermal features. 

Although we benchmark our methodology against specific experiments on graphite, the results are applicable to the broader discussion of electron relaxation and energy redistribution in any optically-excited material system. In addition, this approach is easily adaptable to various pump-probe experiments. For this reason, the code used to simulate our graphite experiment is made available for the simulation of other pump-probe experiments, as well as pedagogical purposes \cite{Code}. 
\section{TR-ARPES on graphite}
\begin{figure*}
    \centering
    \includegraphics{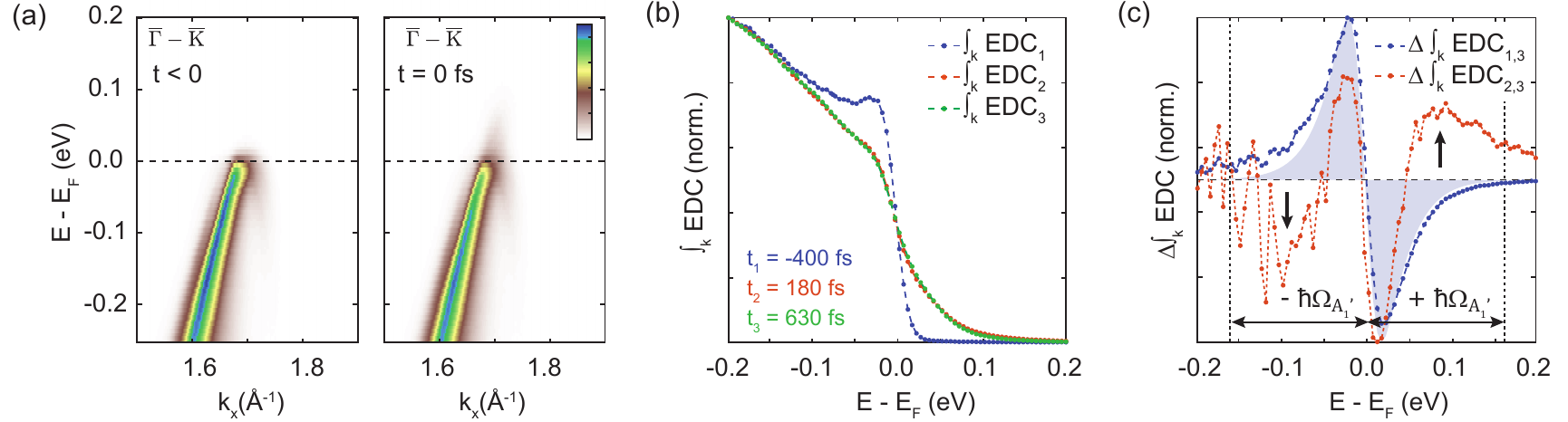}
    \caption{\textbf{TR-ARPES measurements on graphite.} (a) ARPES spectra from graphite measured along the $\overline{\Gamma}$-$\overline{\mathrm{K}}$ direction before pump arrival and at zero-delay. The equilibrium (before pump-arrival) temperature is 50~K. The pump pulse has a photon energy of 1.2~eV and a duration of 120~fs. (b) Momentum-integrated energy distribution curves ($\int_k \mathrm{EDC}$) (linear scale) shows a strong deviation away from the FD function due to the compounded contribution of the dispersion and matrix element effects (see text). (c) Momentum-integrated differential energy distribution curves ($\Delta\int _{k} \mathrm{ EDC}$) computed from the curves in panel (b), using $\int_k \mathrm{EDC}_3$ as a reference. The blue (orange) curve is the $\Delta\int _{k} \mathrm{ EDC}$ at -400~fs (180~fs), and is labelled $\Delta\int_k \mathrm{EDC}_{1,3}$ ($\Delta\int_k \mathrm{EDC}_{2,3}$). The shaded region indicates the difference between two FD distributions, which bears similarity to the blue curve, but cannot describe the orange one. Accumulation of electrons within the phonon window is indicated by black arrows; dashed lines indicate the phonon window associated with the $\mathrm{A_1'}$ optical phonon, with momentum $K$ and energy $\hbar\Omega_{\mathrm{A_1'}}=0.16$~eV.}
    \label{Fig1}
\end{figure*}

We performed TR-ARPES measurements on high quality, single-crystal graphite (details in Appendix \ref{App.Mat}). The pump pulse is the output of a femtosecond ytterbium-doped fiber laser, with 1042~nm fundamental wavelength (1.19~eV). The probe pulse is the 21$^{\mathrm{st}}$ harmonic of the pump (25~eV), produced via high-harmonic generation inside a femtosecond enhancement cavity. The system time and energy resolution is 190~fs and 21~meV \cite{Mills2018}. We chose a low-fluence regime (20~$\mathrm{\mu J}/\mathrm{cm}^2$ incident fluence) in order to emphasize the non-thermal effects \cite{Groeneveld1992}. The negative- and zero-delay (i.e. $t=t_{\mathrm{probe}}-t_{\mathrm{pump}}$) ARPES spectra taken along the $\overline{\Gamma}-\overline{\mathrm{K}}$ direction are shown in Fig.\,\ref{Fig1}(a). The low-energy dispersion of graphite consists of cone-like bands centered at the $\overline{\mathrm{K}}$ and $\overline{\mathrm{K'}}$ points. Away from the  $\overline{\mathrm{K}}$ ($\overline{\mathrm{K'}}$) point, the bands disperse linearly, similar to the Dirac cones of graphene; however, the Dirac fermions are massive, becoming parabolic within $\approx 100$~meV of the Fermi energy ($E_F$) at the $\overline{\mathrm{K}}$ ($\overline{\mathrm{K'}}$) point. As our sample is undoped, we observe occupation of only the valence band up to the crossing point. At zero-delay, we see a small transfer of spectral weight from below to above the $E_F$.

The ARPES intensity can be written as \cite{Damascelli2004}:
\begin{equation}
    I(\mathbf{k},\epsilon)\propto |M_{f,i}^{\mathbf{k}}|^2 A(\mathbf{k},\epsilon) f(\epsilon),
    \label{Eq: ARPES intensity}
\end{equation}
where $|M_{f,i}^{\mathbf{k}}|$ is the matrix element associated with the photoemission process, $A(\mathbf{k},\epsilon)$ is the one-electron removal spectral function, and $f(\epsilon)$ describes the electron occupation. We emphasize that $f(\epsilon)$ is only given by the FD distribution in equilibrium (i.e. no pump, or $t<0$). From inspection of Fig.\,\ref{Fig1}(a), the momentum- and band-dependence of  $|M_{f,i}^{\mathbf{k}}|$ is immediately apparent, with the right branch of the cone almost entirely suppressed. Integrating over momentum, the importance of $|M_{f,i}^{\mathbf{k}}|$ becomes even more pronounced. In the limit of constant $|M_{f,i}^{\mathbf{k}}|$, the momentum-integrated energy-distribution curve ($\int_k \mathrm{EDC}$) is given by $\mathrm{\int_k \mathrm{EDC}}=\int dk_x A(k_x,\epsilon)f(\epsilon)$. This 1D integral corresponds to an occupied tomographic density of states (TDOS) \cite{Reber2012, Boschini2020b}. As such, the $\int_k \mathrm{EDC}$s  should be constant below -0.1~eV due to the linearity of the dispersion and the 1D integration. Above this point, the TDOS increases monotonically towards $E_F$ as the massive Dirac fermion deviates from linearity (Appendix \ref{App.EDCsim}). By contrast, the $\int_k \mathrm{EDC}$s in Fig.\,\ref{Fig1}(b) are not constant in the linear-band regime, indicating significant momentum dependence for $|M_{f,i}^{\mathbf{k}}|$.

Upon excitation, the reduction (increase) in intensity below (above) $E_F$ observed in Fig.\,\ref{Fig1}(b) could conceivably be described as a thermal broadening of the FD distribution, resulting from an increased effective electronic temperature. However, as discussed above, a correct FD fit must account for the aforementioned effects of band dispersion [encoded in $A(\mathbf{k},\omega)$] and photoemission matrix elements. While the computational modelling of these quantities has come impressively far \cite{Gierz2011, Day2019}, the non-thermal features are often subtle enough that it remains challenging to discern whether residuals of the fit should be attributed to model imperfection or a deviation from the FD distribution. Thus, we must show our data to be non-thermal beyond the use of fits. In particular, we will make use of differential momentum-integrated energy-distribution curves ($\Delta\int_k \mathrm{EDC}$s) defined as: 
\begin{equation}
\begin{split}
    \Delta\int _{k} \mathrm{ EDC}(\epsilon, t_1, t_2)=\int dk_x &A(k_x,\epsilon)|M_{f,i}^{k_x}|^2\times\\
    &[f(\epsilon, t_2)-f(\epsilon, t_1)].
\end{split}
\label{Eq. dEDC}
\end{equation}
In principle, both the spectral function $A(\mathrm{k},\omega)$ and the matrix element $|M_{f,i}^{\mathbf{k}}|$ could be time-dependent \cite{Boschini2018, Boschini2020}; however, as the dispersion of graphite undergoes minimal band-renormalization under optical pumping, here we assume the electronic dispersion and matrix element to be constant in time. As a result, the time-dependence is isolated to that of the electronic distribution $f(\epsilon, t)$. Typically, $t_1 < 0$ (before pump arrival), so the reference is characterized by a FD distribution at the initial temperature (here 50~K). However, this reference is not suitable for isolating non-thermal features, which are subtle and obfuscated by the thermal broadening present at all positive delays. Instead, by taking the difference between two different delays with similar electronic temperatures, we can more effectively isolate non-thermal features. To demonstrate this, the $\Delta\int _{k} \mathrm{EDC}$s in Fig.\,\ref{Fig1}(c) uses $\int_k\mathrm{EDC}_3$ at $t_3=630$~fs as a reference, and are defined as $\Delta\int _{k} \mathrm{EDC}_{1,3}$, and $\Delta\int _{k} \mathrm{EDC}_{2,3}$ for delays at $t_1=-400$~fs and $t_2=180$~fs, respectively.

We observe that the blue curve in Fig.\,\ref{Fig1}(c) -- representing $\Delta\int_k \mathrm{EDC}_{1,3}$ -- is positive (negative) below (above) $E_F$, and crosses zero exactly once at $E_F$, consistent with what one expects for thermal broadening (shaded blue region). As we know the unpumped $\int_k\mathrm{EDC}_1$ is surely thermalized, this tells us that the distribution at 630~fs could be, but is not necessarily thermalized. In contrast, the orange curve ($\Delta\int_k \mathrm{EDC}_{2,3}$) shows not one, but three sign changes -- a feature that cannot be affected by photoemission matrix elements or the spectral function, given the definition of Eq.\,\ref{Eq. dEDC}. It also cannot be obtained by taking the difference of FD distributions, which has only one sign change due to the linear energy dependence in the FD exponential. Thus, the three sign changes constitute direct evidence that a non-thermal electronic distribution characterizes $\int_k \mathrm{EDC}_2$ and/or $\int_k \mathrm{EDC}_3$.

Finally, we note that $\Delta\int _{k} \mathrm{EDC}_{2,3}$ shows an accumulation of electrons and holes within a region $\pm 0.16$~eV around the Fermi level [see arrows in Fig.\,\ref{Fig1}(c)]. In graphite, strongly-coupled optical phonons with momentum $K$ and energy $\hbar\Omega_{\mathrm{A_1'}}=0.16~$eV constitute a major channel through which electrons relax \cite{Na2019}; however, close to $E_F$, this channel is frozen out, as the final states (below $E_F$) are already occupied. The bottle-necking of this relaxation channel creates an accumulation of electrons exactly between the energies $\pm\hbar\Omega_{\mathrm{A_1'}}$, indicated by dashed lines in Fig.\,\ref{Fig1}(c). This prompts us to label the region $[-\hbar\Omega_{\mathrm{A_1'}}, \hbar\Omega_{\mathrm{A_1'}}]$ as the ``phonon window", defined by a given phonon energy $\hbar\Omega_{\mathrm{A_1'}}$ \cite{Sentef2013, Kemper2018}.  

\begin{figure*}[t!]
    \centering
    \includegraphics{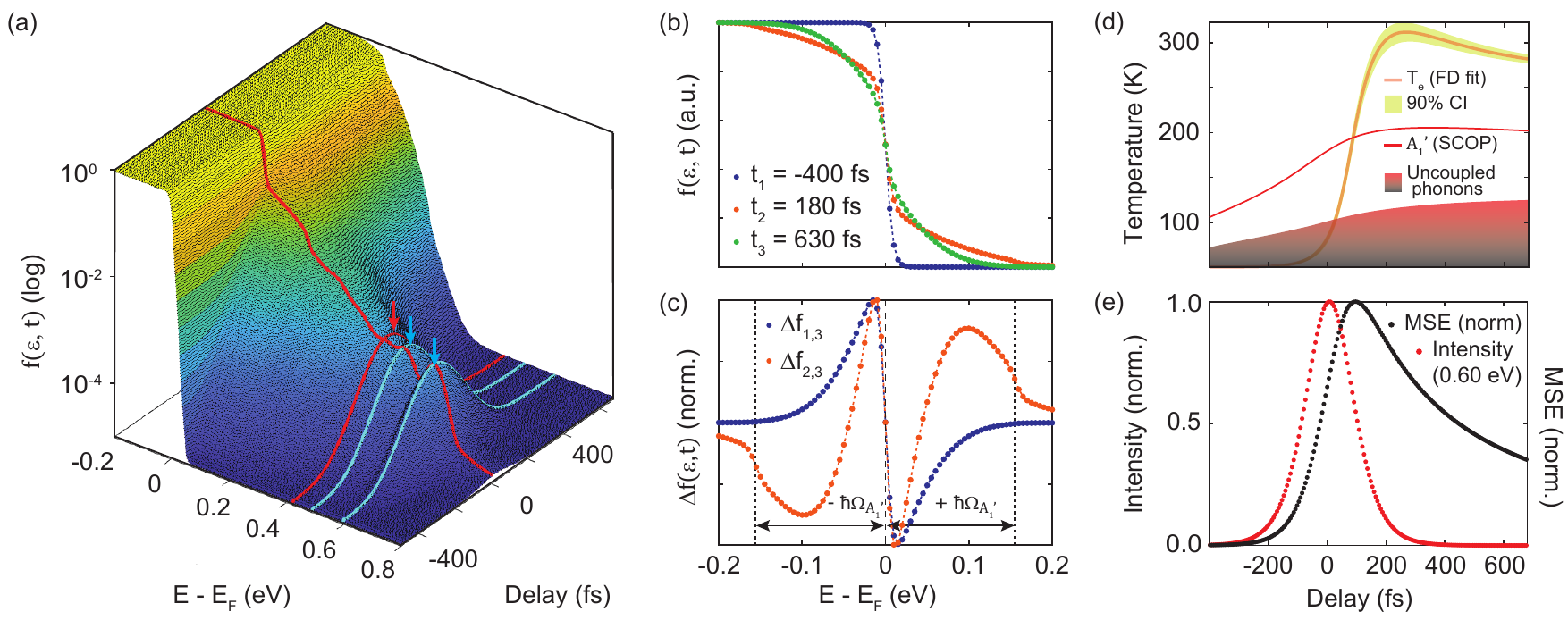}
    \caption{\textbf{Boltzmann simulations of electron relaxation in graphite.} (a) The occupation function $f(\epsilon, t)$ calculated from the Boltzmann rate equations (log scale). The zero-delay distribution is given by the red slice; two peaks associated with optical transitions and a peak created through scattering with the strongly-coupled optical phonon (SCOP) are highlighted in blue and red, respectively. The dynamics of each peak are also highlighted. For visualization, we add a background of $10^{-5}$ to $f(\epsilon, t)$ to represent the dynamic range of the experiment. (b) $f(\epsilon, t)$ at the same delays in Fig.\,\ref{Fig1}(b). (c) Differential occupation using $f(\epsilon, t_3)$ as a reference. $\Delta f_{1,3}$ and $\Delta f_{2,3}$ are shown in blue and orange, respectively, reproducing the changes in sign observed in the data [Fig.\,\ref{Fig1}(c)]. The phonon window associated with the $\mathrm{A_1'}$ mode is indicated by black dashed lines. (d) Electron and phonon effective temperatures extracted by fitting the distribution with a FD and inverting the Bose-Einstein distribution, respectively. Yellow shaded region indicates the 90\% confidence interval of the fit. Optical (high energy) phonons are shaded in red; acoustic phonons are shaded in black. (e) The occupation in a 0.1~eV window centered at 0.6~eV is given by the red markers. The mean-squared error [$\mathrm{MSE}=1/n\sum_n(y_{\mathrm{fit}}-y)^2$] of the fit is given by black markers. MSE is a quantitative measure of the non-thermal contributions to the electronic distribution, and it is non-zero long after the pump pulse has passed.}
    \label{Fig2}
\end{figure*}

\section{Boltzmann model}
To determine when a temperature-based model becomes appropriate, we simulate the electron occupation function $f(\epsilon, t)$ using Boltzmann rate equations, and compare it to the FD distribution. The temporal evolution of $f(\epsilon, t)$ and the phonon occupation function $n(\Omega ,t)$ is governed by a series of coupled rate equations \cite{Rethfeld2002, Ono2018}:
\begin{equation}
    \begin{split}
        \frac{\partial f}{\partial t}&=V\left(\frac{\partial f}{\partial t}\right)_{\mathrm{e-e}}+G\left(\frac{\partial f}{\partial t}\right)_{\mathrm{e-ph}}+\Phi\left(\frac{\partial f}{\partial t}\right)_{\mathrm{inj}}\\
        \frac{\partial n}{\partial t}&=G\left(\frac{\partial n}{\partial t}\right)_{\mathrm{ph-e}}+A\left(\frac{\partial n}{\partial t}\right)_{\mathrm{ph-ph}}.\\
    \end{split}
\label{Eq2}
\end{equation}
The photo-excitation of electrons is described by the injection term $\left(\partial f/\partial t\right)_{\mathrm{inj}}$. As each photon creates one electron-hole pair, the functional form of the injection rate follows the time-domain pulse shape. The number of photo-excited electrons for each resonant excitation is given by the fluence and Fermi's golden rule, which we combine into a single constant $\Phi$. Following the photo-excitation, the electronic distribution is determined chiefly by the interplay between e-e and e-ph scattering, described by $\left(\partial f/\partial t\right)_{\mathrm{e-e}}$ and $\left(\partial f/\partial t\right)_{\mathrm{e-ph}}$ respectively. For simplicity, we consider coupling to only the strongly-coupled $\mathrm{A_1'}$ phonon mode in $\left(\partial f/\partial t\right)_{\mathrm{e-ph}}$. The scattering rate of e-e and e-ph is given by the constants $V$ and $G$, respectively. From the phonon perspective, e-ph coupling increases the phonon occupation, as energy is transferred from the electron bath to the phonon bath. This is described in the term $\left(\partial n/\partial t\right)_{\mathrm{ph-e}}$, and mediated by the e-ph coupling $G$. As the occupation of the $\mathrm{A_1'}$ mode increases, it anharmonically decays into lower-energy modes. We capture this in the ph-ph scattering term $\left(\partial n/\partial t\right)_{\mathrm{ph-ph}}$, the rate of which is given by $A$. Details of the calculation and the functional form of each term are given in Appendix \ref{App.Derivation} and \ref{App.Sim}. 

The resultant occupation function in log scale is shown in Fig.\,\ref{Fig2}(a). We highlight $f(\epsilon,t=0)$ in red, the blue arrows indicate peaks arising from the optical injection, while the red arrow indicates the phonon-induced replica, created by electrons scattering with the $\mathrm{A_1'}$ phonon from the state at 0.6~eV to the state at 0.44~eV \cite{Na2019}. The simulated occupation function at delays corresponding to Fig.\,\ref{Fig1}(b) are displayed in Fig.\,\ref{Fig2}(b), and the differential curves corresponding to those in Fig.\,\ref{Fig1}(c) are also shown in Fig.\,\ref{Fig2}(c). From these figures, we see that our simulations reproduce the changes in sign and the accumulation of electrons within the phonon window. 
\begin{figure*}[t!]
    \centering
    \includegraphics{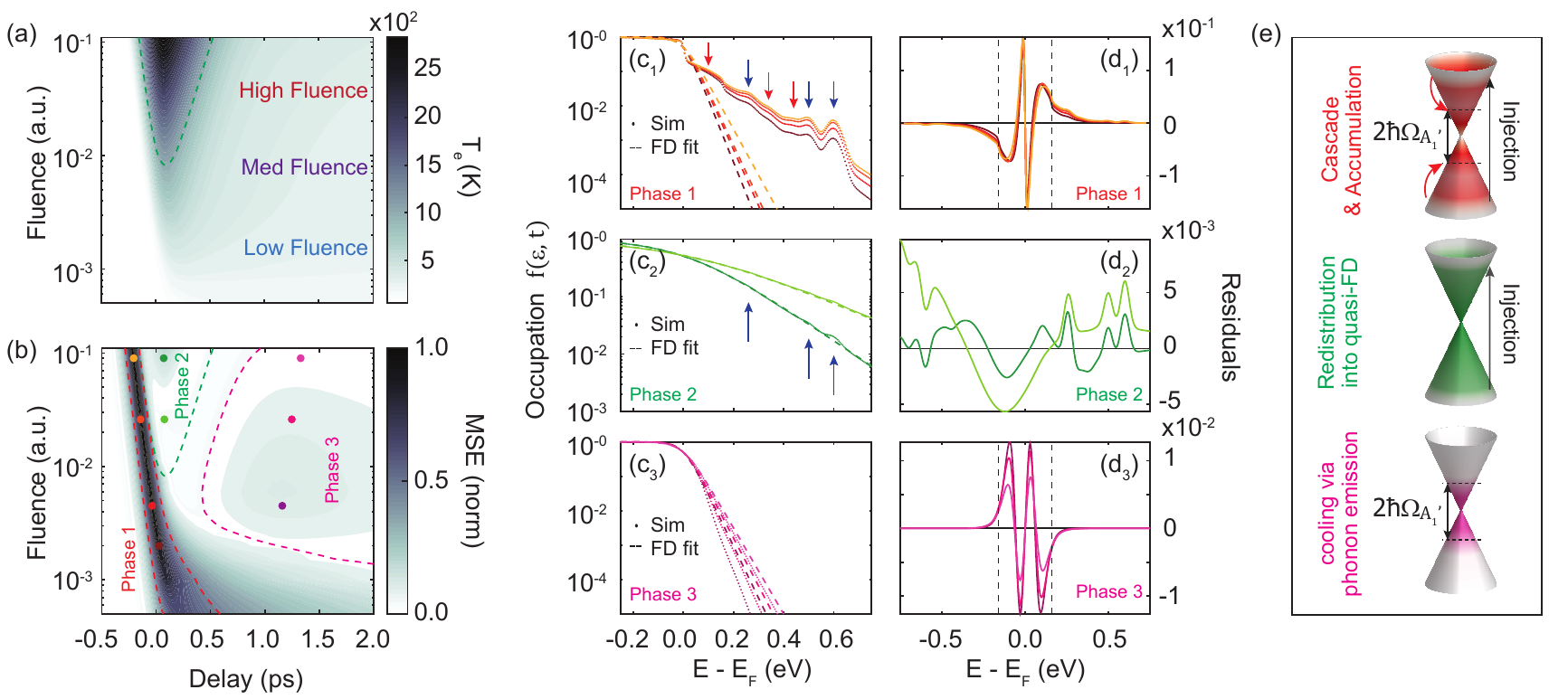}
    \caption{\textbf{Non-thermal phases in the fluence-delay phase space.} (a) Effective electronic temperature ($T_e$) extracted from a FD fit of the occupation function. The green dashed line is a contour at 1000~K; the different fluence regimes are qualitatively indicated by labels. (b) MSE obtained from the fit of the effective electronic temperature. Three phases are outlined by red, green, and magenta dashed lines, respectively; colored markers indicate the delay and fluence at which the occupation function and residuals are shown in panels (c) and (d). (c$_{1-3}$) Simulated distribution $f(\epsilon, t)$ in Phase 1, 2, and 3, respectively, are given by dotted lines matching the colored markers in panel (b). FD fits of the occupation function are shown in dashed lines in corresponding colors. The blue and red arrows in Phase 1 and 2 highlight the peaks corresponding to optical excitation and phonon-emission, respectively. (d$_{1-3}$) Residuals of the FD fit at delays corresponding to the markers in panel (b); despite being taken at different delays, residuals in the same phase have the same qualitative features, supporting the classification into these separate phases. (e) Cartoons indicating the primary mechanisms in the three identified phases: (Phase 1) injection, cascade, and accumulation within the phonon window; (Phase 2) redistribution of accumulated electrons into an energetic quasi-FD distribution; (Phase 3) cooling of the distribution inside the phonon window via phonon emission.}
    \label{Fig3}
\end{figure*}

We also note that differential curves are not the optimal way to visualize the non-thermal electron distribution because the reference (here $\int\mathrm{EDC}_3$) is not necessarily thermal. However, in the simulation, we can access the occupation function $f(\epsilon, t)$, unencumbered by details of the dispersion, matrix elements, and experimental resolution present in ARPES intensity. Therefore, the non-thermal electrons are directly accounted for in the residuals of a FD fit of the simulated $f(\epsilon, t)$. We study the effective electron temperature $T_e$ extracted from such a fit in Fig.\,\ref{Fig2}(d). The small $\Phi$ used to match the low fluence used in the experiment manifests as a moderate transient increase of $T_e$ followed by slow decay. From $n(\Omega,t)$, we can also access the phonon ``temperature", which is obtained by inverting $n(\Omega ,t)$ using the Bose-Einstein distribution for each $\Omega$. Remarkably, although the pump pulse injects energy into the electron bath, we observe that the occupation temperature of the SCOP increases faster, and is hotter than the electronic temperature. To a lesser extent, this is also true for the uncoupled modes (shaded region), which increase in occupation via the anharmonic decay of the SCOP. This result is starkly different from the view presented by TTM -- in which the electronic bath heats to a high temperature before transferring energy to the phonon bath -- further highlighting the inappropriate definition of temperature in this transient regime. 

We quantify the contribution of non-thermal electrons by using the mean-squared-error (MSE) of the fit, defined as $\mathrm{MSE}=1/n\sum_n(y_{\mathrm{fit}}-y)^2$, and shown as black markers in Fig.\,\ref{Fig2}(e). We compare MSE to the intensity at 0.6~eV (where electrons are optically injected), shown in Fig.\,\ref{Fig2}(e) in red markers. The latter closely follows the excitation pulse shape in time; however, the non-thermal electron contribution is substantial even at 600~fs. We remark that MSE can also be extracted from fits of the data with a phenomenological model -- which we discuss in the supplementary -- though contributions by the spectral function and matrix elements complicates its interpretation (Appendix \ref{App.Comp}). In the following, we use MSE from the simulation to classify different regimes of non-thermal occupation, for a range of fluences.

\section{The Non-thermal Window}
The motivation for looking at the fluence-dependence of MSE is twofold. First, non-thermal features related to e-ph scattering are expected to become more important at low fluences and low initial temperatures \cite{Groeneveld1992, Ishida2011a}. While we have seen non-thermal occupation of electrons in graphite in low-fluence experiments, high-fluence experiments on graphite and graphene report a thermalized distribution within the time-resolution of those experiments ($\approx 10$~fs) \cite{Gilbertson2012, Johannsen2013, Gierz2013c, Stange2015, Yang2016}. Thus, we want to understand the evolution of the non-thermal distribution in the fluence-delay phase space. Secondly, fluence plays a crucial role in designing pump-probe experiments and in determining the relevant physics encoded. Given the excellent agreement between our simulation and data at low fluence, we keep the scattering strengths $V$, $G$, and $A$ unchanged, while varying the fluence $\Phi$. 

As before, we study the evolution of the electronic temperature extracted by a FD fit [Fig.\,\ref{Fig3}(a)]. Although the fluence $\Phi$ is in arbitrary units, we can make quantitative comparisons between simulation and experiment based on the extracted electronic temperature. Here, we qualitatively characterize the low-fluence regime as those distributions with a peak electronic temperature $T_e<500$~K, and the high-fluence regime for $T_e>1000$~K. The MSE as a function of fluence and pump-probe delay are presented in Fig.\,\ref{Fig3}(b). We reiterate that, as we fit the electron occupation with a FD distribution, the residuals of the fit arise purely from non-thermal occupation. Thus MSE provides a quantitative measure of non-thermal electrons in the electronic distribution. Three phases are apparent upon observation of the fluence-delay phase space, enclosed by red, green, and magenta dashed lines, respectively, in Fig.\,\ref{Fig3}(b). We compare the simulated $f(\epsilon,t)$ and the FD fit for the three phases in Fig.\,\ref{Fig3}(c$_{1}$-c$_{3}$), respectively, with corresponding residuals shown in Fig.\,\ref{Fig3}(d$_{1}$-d$_{3}$). The illustration in Fig.\,\ref{Fig3}(e) describes the physical processes underlying each phase.

We begin the discussion of the different phases in the low-fluence regime, which matches conditions explored in Figs.\,\ref{Fig1} and \ref{Fig2}. In this regime, Phase 1 (red dashed lines) is long-lasting, extending to hundreds of femtoseconds. The occupation function in this phase is shown in Fig.\,\ref{Fig3}(c$_1$) at several fluences indicated by markers in Fig.\,\ref{Fig3}(b). We observe two distinct slopes, on top of which sits a series of peaks: electrons are photo-excited into states shown by dashed blue arrows, and subsequent peaks are created via the emission of $\mathrm{A_1'}$ optical phonons. The smooth exponential background below the peaks is given by e-e scattering. The peaks constitute a small part of the non-thermal features; in fact, the most prominent non-thermal feature here is the accumulation of electrons within the phonon window, evident in the residuals [Fig.\,\ref{Fig3}(d$_1$)] and whose phenomenology is illustrated in the top sketch of Fig.\,\ref{Fig3}(e) (dashed lines indicate the edges of the phonon window). This accumulation is the same feature observed in Fig.\,\ref{Fig1}(c) and \ref{Fig2}(c), though here the FD is used as a reference. As electrons relax, the distribution moves directly towards Phase 3, which is the longest-lasting phase. In this phase, we no longer observe injected carriers in the occupation function [Fig.\,\ref{Fig3}(c$_{3}$)]: electrons accumulated inside the phonon window cool slowly via emission of $A_1'$ phonons, while e-e scattering redistributes the electrons in response to the reduced energy in the electron system [bottom sketch of Fig.\,\ref{Fig3}(e)]. The amplitude of the peaks in Fig.\,\ref{Fig3}(d$_{3}$) diminishes as energy is lost until the distribution reaches equilibrium. Note that we have not included dissipation in the Boltzmann equations; hence energy cannot leave the e-ph system, the final temperature will be higher than the initial temperature, and the dynamics will be inaccurate on timescales of energy dissipation from the area illuminated by the pump pulse (here nominally 200~$\mathrm{\mu m}\times$ 400~$\mathrm{\mu m}$)\cite{Jago2019}. 

As the fluence increases, Phase 1 begins at earlier delays and ends much quicker, even at the rising edge of the excitation pulse (negative delays). We then reach Phase 2, where the density of hot electrons dramatically increases, limiting the available phase space for electron relaxation. This phase boundary can be defined as a crossover between classical and quantum statistics, which occurs when the electron density condition $f(\epsilon, T)\ll 1$ is no longer satisfied. Applying this criterion to the middle of the photo-excited electron distribution (approximately 0.4~eV), we find that the crossover occurs at the effective temperature $T_e\simeq 1000$~K. Thus, the boundary of Phase 2 is delimited by a constant temperature contour at 1000~K in Fig.\,\ref{Fig3}(a) and (b). Below this effective temperature, the phase space is largely empty, allowing electrons to occupy states unevenly, i.e. in a non-thermal way. Above this effective temperature, electrons strongly feel Pauli-blocking -- induced by increased electron density -- and redistribute according to fermion statistics, giving rise to a quasi-FD distribution [see Fig.\,\ref{Fig3}(c$_2$)]. The residuals in this phase also have the highest variance. While we still observe some identifiable features related to photo-excitation [Fig.\,\ref{Fig3}(d$_2$)], the intensity of the features is low and easily lost in the noise of real measurements. Following this, the system moves into Phase 3, albeit at a higher temperature than that seen in low-fluence experiments. 

In comparing the low and high fluence regimes, we note that the low fluence Phase 1 and high-fluence Phase 2 occurs at approximately the same pump-probe delay. The timescales of these two phases match both our observation of non-thermal features near zero-delay (see Fig.\,\ref{Fig1}), as well as the observations of a (nearly) well-thermalized distribution within the time resolution for high fluence studies \cite{Gilbertson2012, Johannsen2013, Gierz2013c, Stange2015, Yang2016}. Lastly, we note that even while a FD can fit the electron bath in Phases 2 and 3, the phonon bath may not be thermalized for many more picoseconds [as previously seen in Fig.\,\ref{Fig2}(d)], violating the assumptions of the TTM. In general, the non-thermal/thermal boundary is an intrinsic property of each material system and dependent on experimental conditions, and should be determined with careful analysis. The different non-thermal regimes will be determined by which scattering processes are available during relaxation, their timescales and corresponding bottlenecks, as well as the density of electrons in different regions of the bandstructure.

\section{Conclusion}
In this work, we have demonstrated non-thermal features near the Fermi-level in time-resolved pump-probe ARPES experiments. In particular, we have used Boltzmann rate equations to simulate the excitation and relaxation of electrons and phonons in graphite, using low-fluence TR-ARPES data to benchmark our momentum-averaged coupling constants. We illustrated the shortcomings of temperature-based approaches at low fluence by applying the temperature analysis to a simulated occupation function. By separating time-dependent residuals from the purely thermal electron distribution, we identified three different non-thermal phases, spanning several hundreds of femtoseconds in the fluence-delay phase space. The residuals of the FD fit here directly constitute non-thermal contributions to the electron distribution function and showcase the distinct processes taking place in each regime. We account for the apparent disparity in the observation of non-thermal features across low and high-fluence experiments in terms of a crossover between the classical and quantum regimes of electron density. 

Our model also captures some salient aspects of fully quantum mechanical approaches such as the non-equilibrium Keldysh formalism, which encodes particle properties in propagators \cite{Abdurazakov2018, Kemper2018, Omadillo}. While these approaches are compelling, the calculations are complex and computationally taxing. We also note that the discussion here has been confined to relaxation processes in graphite for a specific set of experimental parameters. For these conditions, we reliably observed the three non-thermal phases in our exploration of the parameter space for the Boltzmann model; however, other materials and experimental parameters might manifest these phases differently, or host new phases altogether. That said, the rate-equation model is easily adaptable to other material systems by changing the density of states and the momentum averaged coupling constants in the rate equations. In conclusion, the Boltzmann approach offers an instructive and intuitive way to understand non-equilibrium processes without leaning on thermodynamic variables. One can also use these models pedagogically, by simulating electron relaxation in parabolic and linear dispersions or -- by toggling e-e, e-ph, or ph-ph scattering on and off -- to see the corresponding effects on the evolution of the electronic distribution. To this end, the code used for this paper has been made available in \cite{Code}. 

\section{Acknowledgements}
We gratefully acknowledge H. H. Kung, C. Guti\'errez, at the Quantum Matter Institute, G. Chiriaco, A. J. Millis and A. Georges at the Flatiron Institute, and T. P. Devereaux at Stanford University, for helpful discussions. This research was undertaken thanks in part to funding from the Max Planck-UBC-UTokyo Centre for Quantum Materials and the Canada First Research Excellence Fund, Quantum Materials and Future Technologies Program. This project is also funded by the Gordon and Betty Moore Foundation's EPiQS Initiative, Grant GBMF4779 to A.D. and D.J.J.; Killam, Alfred P. Sloan, and Natural Sciences and Engineering Research Council of Canada’s (NSERC’s) Steacie Memorial Fellowships (A.D.); the Alexander von Humboldt Fellowship (A.D.); the Canada Research Chairs Program (A.D.); NSERC, Canada Foundation for Innovation (CFI); British Columbia Knowledge Development Fund (BCKDF); and the CIFAR Quantum Materials Program. A. F. K. acknowledges support by the National Science Foundation under grant DMR-1752713.
\onecolumngrid
\appendix
\section{Materials and Methods}
\label{App.Mat}
We measure naturally occurring high-quality graphite. The samples were cleaved in-situ at a base pressure of $5\times10^{-11}$ Torr. The pump-induced average heating was determined to be 50~K from the Fermi-edge broadening. We use the Scienta R4000 hemispherical analyzer for photoelectron detection, with an energy resolution of $< 2$~meV. 

The pump pulse is the output of a Yb-fiber laser operating at 60~MHz, with 1.19~eV photon energy, 120~fs pulse duration, and 30~meV bandwidth. The probe pulse is produced via high-harmonic generation (HHG) at the focus of a femtosecond enhancement cavity (fsEC) using a krypton gas jet. We out-couple the harmonics with a grating mirror and the select the 21st harmonic (25~eV) for photoemission. The pulse duration and bandwidth of the probe are 22~meV and 150~fs, respectively. For this experiment, the pump and probe both have s-polarization. The spot size on the sample is $200~\mu m \times 400\mu m$. Details of the laser are documented elsewhere \cite{Mills2018}.
\begin{figure}[t!]
    \centering
    \includegraphics{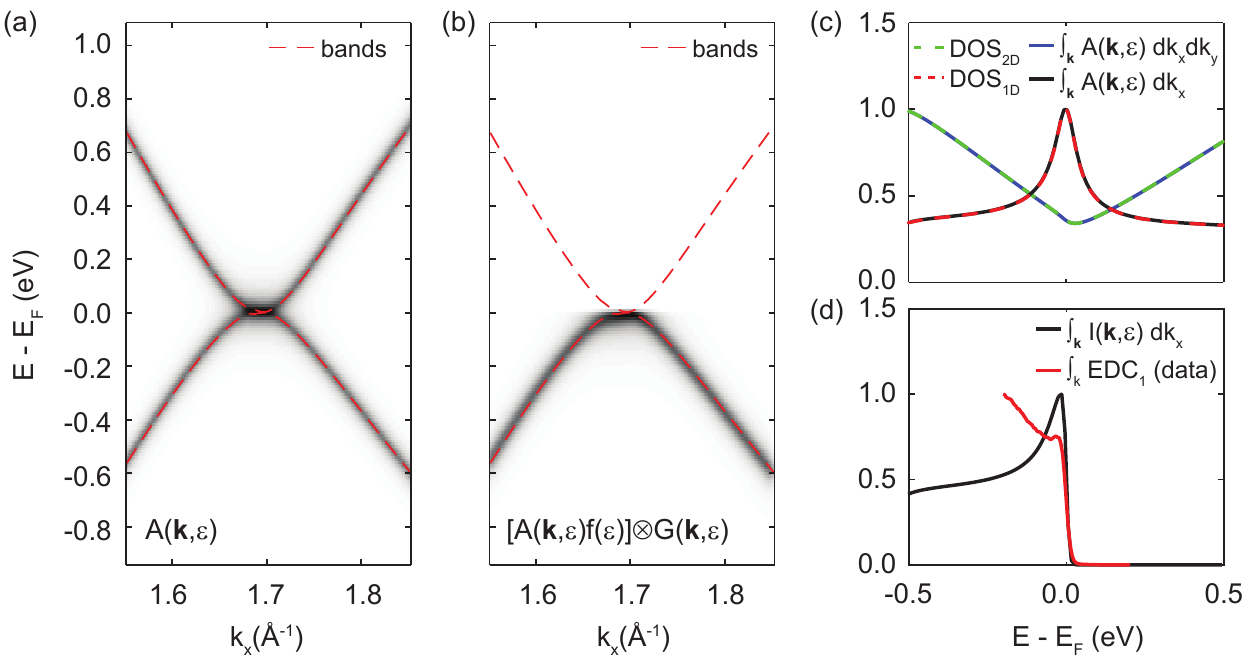}
    \caption{\textbf{Simulation of the $\int_k\mathrm{EDC}$}. (a) The spectral function calculated with $\Sigma'=0$, $\Sigma''=30$~meV, along the $\Gamma - \mathrm{K}$ direction for $k_z=0.2$~\AA$^{-1}$. The dispersion $\epsilon_{\mathbf{k}}$ is shown in dashed red lines. (b) The spectral function modified by the Fermi-Dirac, and convolved with a Gaussian in energy and momentum to account for system resolution effects. (c) Intensity profiles obtained by integrating the spectral function over $dk_x$, at $k_y=0$ (black) and $dk_x dk_y$ (blue). Corresponding density of states are given in red and green dashed lines, respectively. (d) Intensity profiles obtained by integrating the modified spectral function over $dk_x$, at $k_y=0$ (black) and $dk_xdk_y$ (blue).}
    \label{SFig1}
\end{figure}
\section{Simulating the momentum-integrated \\ energy distribution curve}
\label{App.EDCsim}
In this section, we discuss the effect of the spectral function $A(\mathbf{k},\epsilon)$ and the photoemission matrix element on the momentum-integrated EDC ($\int_k \mathrm{EDC}$) presented in Fig.\,1(b) of the main text. The spectral function is calculated from the following equation \cite{Damascelli2004}:
\begin{equation}
A(\mathbf{k},\epsilon)=\frac{\Sigma''}{(\epsilon-\epsilon_{\mathbf{k}}-\Sigma')^2+(\Sigma'')^2}.
\end{equation}
For simplicity, we set the real part of the self-energy ($\Sigma'$) to zero and set the imaginary part of the self energy to a finite value ( $\Sigma''=30$~meV) for visualization purposes. The dispersion $\epsilon_\mathbf{k}$ is calculated using the Wien2K \cite{w2k}. 
In Fig.\,\ref{SFig1} (a), we show the spectral function along the $\Gamma -K$ direction, at $k_z=0.2$~\AA$^{-1}$, similar to our experiment. The integration of the spectral function in one and two dimensions (i.e. over $dk_x$ at $k_y=0$ and $dk_xdk_y$, respectively) is given by blue and black solid lines in Fig.\,\ref{SFig1}(c). These intensity profiles exactly overlap with the density of states (DOS) calculated for the same regions of momentum space. The one-dimensional DOS is nearly flat above 0.2~eV, and peaks towards the crossover of the bands at $E_F$. This results from the 1-dimensionality of the integration and the quasi-linear dispersion above 0.2~eV, which only becomes quadratic as it approaches $E_F$. As well, in the two-dimensional DOS we see the characteristic `V' shape of linearly dispersing bands in 2D. The two bands have slightly different effective mass, as can be seen in the DOS$_\mathrm{2D}$. At $E_F$, the DOS$_\mathrm{2D}$ takes on finite value as the dispersion becomes quadratic. 

Next, we modify the intensity profile into what we would see in the experiment (neglecting matrix elements). The modified intensity is given by:
\begin{equation}
    I(\mathbf{k},\epsilon)=[A(\mathbf{k},\epsilon)f(\epsilon)]*G(\mathbf{k},\epsilon),
\end{equation} 
where $f(\epsilon)$ is the occupation function (a Fermi-Dirac in equilibrium), and $G(\mathbf{k},\epsilon)$ is a Gaussian in energy and momentum that accounts for the experimental resolution. The modified intensity map is shown in Fig.\,\ref{SFig1}(b), and the integrated intensity profile is shown in Fig.\,\ref{SFig1}(d). As the $\int_k \mathrm{EDC}$ is obtained from a one-dimensional integration (along $k_x$, at $k_y=0$), in the analysis, we use the one-dimensional intensity profile for comparison. In the data, we see a decrease in intensity followed by a small peak; in the simulation, we observe a quasi-linear profile peaking at $E_F$. The photoemission matrix elements are responsible for the difference between these profiles.
\section{Boltzmann rate-equation model: Derivation}
\label{App.Derivation}
In this section, we discuss the derivation of each term of Eq.\,2 in the main text. \\
\begin{equation}
    \begin{split}
        \frac{\partial f}{\partial t}&=\left(\frac{\partial f}{\partial t}\right)_{\mathrm{e-e}}+\left(\frac{\partial f}{\partial t}\right)_{\mathrm{e-ph}}+\left(\frac{\partial f}{\partial t}\right)_{\mathrm{inj}}\\
        \frac{\partial n}{\partial t}&=\left(\frac{\partial n}{\partial t}\right)_{\mathrm{ph-e}}+\left(\frac{\partial n}{\partial t}\right)_{\mathrm{ph-ph}}.\\
    \end{split}
\end{equation}
\textbf{Photo-excitation:} The photo-excitation of electrons consists of depletion of a state $\epsilon'_{\mathbf{k'}}$ below $E_F$ and a population of a state $\epsilon_{\mathbf{k}}$ above $E_F$, such that $h\nu=\epsilon-\epsilon'$. Since the momentum contributed by the photon is negligible, $k=k'$. Each photon creates an electron-hole pair; the injection rate follows the shape of the pump pulse:
\begin{equation}
\left(\frac{\partial f(\epsilon)}{\partial t}\right)_{\mathrm{inj}}=S(\epsilon,\sigma_{\epsilon})T(t,\sigma_t).
\label{Eq: Carrier_inj}
\end{equation}
Here, $T(t,\sigma_t)$ is a generic pulse shape in the time domain with a characteristic full-width at half-maximum (FWHM) $\sigma_t$.  $S(\epsilon, \sigma_{\epsilon})$ determines the states depleted/populated according to the bandwidth of the pump pulse and the optical-joint density of states (OJDOS). As an example, a Gaussian pulse with FWHM $\sigma_\epsilon$ is given by the following $S(\epsilon, \sigma_\epsilon)$:
\begin{equation}
        S(\epsilon,\sigma_{\epsilon})=\sum_i\Phi_i^{\mathrm{inj}}\exp\left(-\frac{(\epsilon-\epsilon^{\mathrm{pop}}_i)^2}{2\sigma_\epsilon^2}\right)-\Phi_i^{\mathrm{dep}}\exp\left(-\frac{(\epsilon-\epsilon^{\mathrm{dep}}_i)^2}{2\sigma_\epsilon^2}\right).
\end{equation}
Here the sum over $i$ includes each optical transition in the OJDOS. $\Phi_i^{\mathrm{pop}}$ and $\Phi_i^{\mathrm{dep}}$ satisfy the relation $\Phi_i^{\mathrm{pop}}N(\epsilon_i^{\mathrm{pop}})=\Phi_i^{\mathrm{dep}}N(\epsilon_i^{\mathrm{dep}})$ -- where $N(\epsilon)$ is the electron density of states -- such that the total number of electrons is conserved. \\
\textbf{Electron-electron scattering:} The electron-electron (e-e) scattering term is given by \cite{Kabanov2008}:
\begin{equation}
\begin{split}
\frac{\partial f_{\mathbf{k}}}{\partial t}=\frac{2\pi}{\hbar}&\sum_{\mathbf{p},\mathbf{q}}V^2_c(\mathbf{q})
\delta(\xi_{\mathbf{k}}-\xi_{\mathbf{k}+\mathbf{q}}+\xi_{\mathbf{p}}-\xi_{\mathbf{p}-\mathbf{q}})\\
&\times[f_{\mathbf{k}+\mathbf{q}}f_{\mathbf{p}-\mathbf{q}}(1-f_{\mathbf{k}})(1-f_{\mathbf{p}})-f_{\mathbf{k}}f_{\mathbf{p}}(1-f_{\mathbf{k}+\mathbf{q}})(1-f_{\mathbf{p}-\mathbf{q}})].\\
\end{split}
\end{equation}
Here $V_c(\mathbf{q})$ is the scattering pseudo-potential, and $\xi_{\mathbf{k}}$ is the electron energy with respect to the equilibrium chemical potential The first term in the equation describes the scattering of an electron from the state $\xi_{\mathbf{k}+\mathbf{q}}$ into the state $\xi_{\mathbf{k}}$. A second electron scatters from the state $\xi_{\mathbf{p}-\mathbf{q}}$ into state $\xi_{\mathbf{p}}$ such that the energy of the electronic system is conserved. Similarly, the second term is the probability of scattering out of the state $\xi_{\mathbf{k}}$.

Averaging over all momenta $\mathbf{k}$:
\begin{equation}
\frac{\partial f(\xi)}{\partial t}\equiv\frac{1}{N(\xi)}\sum_{\mathbf{k}}\delta(\xi_{\mathbf{k}}-\xi)\frac{\partial f_{\mathbf{k}}}{\partial t},
\end{equation}
where $N(\xi)$ is the density of states (DOS). The momentum averaged equation is then \cite{Kabanov2008}:
\begin{equation}
\begin{split}
\frac{\partial f(\xi)}{\partial t}=&\int d\xi'\int d\epsilon\int \epsilon'K(\xi,\xi',\epsilon,\epsilon')\delta(\xi-\xi'+\epsilon-\epsilon')\\
&\times[f(\xi')f(\epsilon')(1-f(\xi))(1-f(\epsilon))-f(\xi)f(\epsilon)(1-f(\xi'))(1-f(\epsilon'))]
\end{split}
\label{Eq: ee k to E}
\end{equation}
where
\begin{equation}
K(\xi,\xi',\epsilon,\epsilon')=\frac{2\pi}{\hbar}\frac{1}{N(\xi)}\sum_{\mathbf{k},\mathbf{p},\mathbf{q}}V^2_c(\mathbf{q})
\delta(\xi_{\mathbf{k}}-\xi)\delta(\xi_{\mathbf{k}+\mathbf{q}}-\xi')\delta(\xi_{\mathbf{p}}-\epsilon)\delta(\xi_{\mathbf{p}-\mathbf{q}}-\epsilon').
\end{equation}
Here the kernel $K$ includes all of the averaging over momenta, while $\delta(\xi-\xi'+\epsilon-\epsilon')$ specifies the elastic scattering condition. It is common to approximate $V_c(\mathbf{q})=V_c$, since the kernel changes on the scale of the Fermi energy $E_F$. We know that the DOS (in units of 1/eV) is defined as:
\begin{equation}
N(\xi)=\sum_{\mathbf{k}}\delta(\xi_\mathbf{k}-\xi),
\end{equation}
So the kernel can be expressed in terms of the DOS:
\begin{equation}
K(\xi,\xi',\epsilon,\epsilon')=\frac{2\pi}{\hbar}\frac{V_c^2}{N(\xi)}[N(\xi)N(\xi')N(\epsilon)N(\epsilon')].
\end{equation}
We put the constants together and relabel it as  $V$, and define $\eta=\xi-\xi'$, such that $\epsilon'=\epsilon+\eta$, and $\xi'=\xi-\eta$. Lastly, we use the Dirac-delta to get rid of one of the integrals, so that Eq.\,\ref{Eq: ee k to E} becomes:
\begin{equation}
\begin{split}
\frac{\partial f(\xi)}{\partial t}=&\frac{V}{N(\xi)}\int d\eta\int d\epsilon[N(\xi)N(\xi-\eta)N(\epsilon)N(\epsilon+\eta)]\\
\times&[f(\xi-\eta)f(\epsilon+\eta)(1-f(\xi))(1-f(\epsilon))-f(\xi)f(\epsilon)(1-f(\xi-\eta))(1-f(\epsilon+\eta))].
\end{split}
\end{equation}
\begin{figure}[t!]
    \centering
    \includegraphics{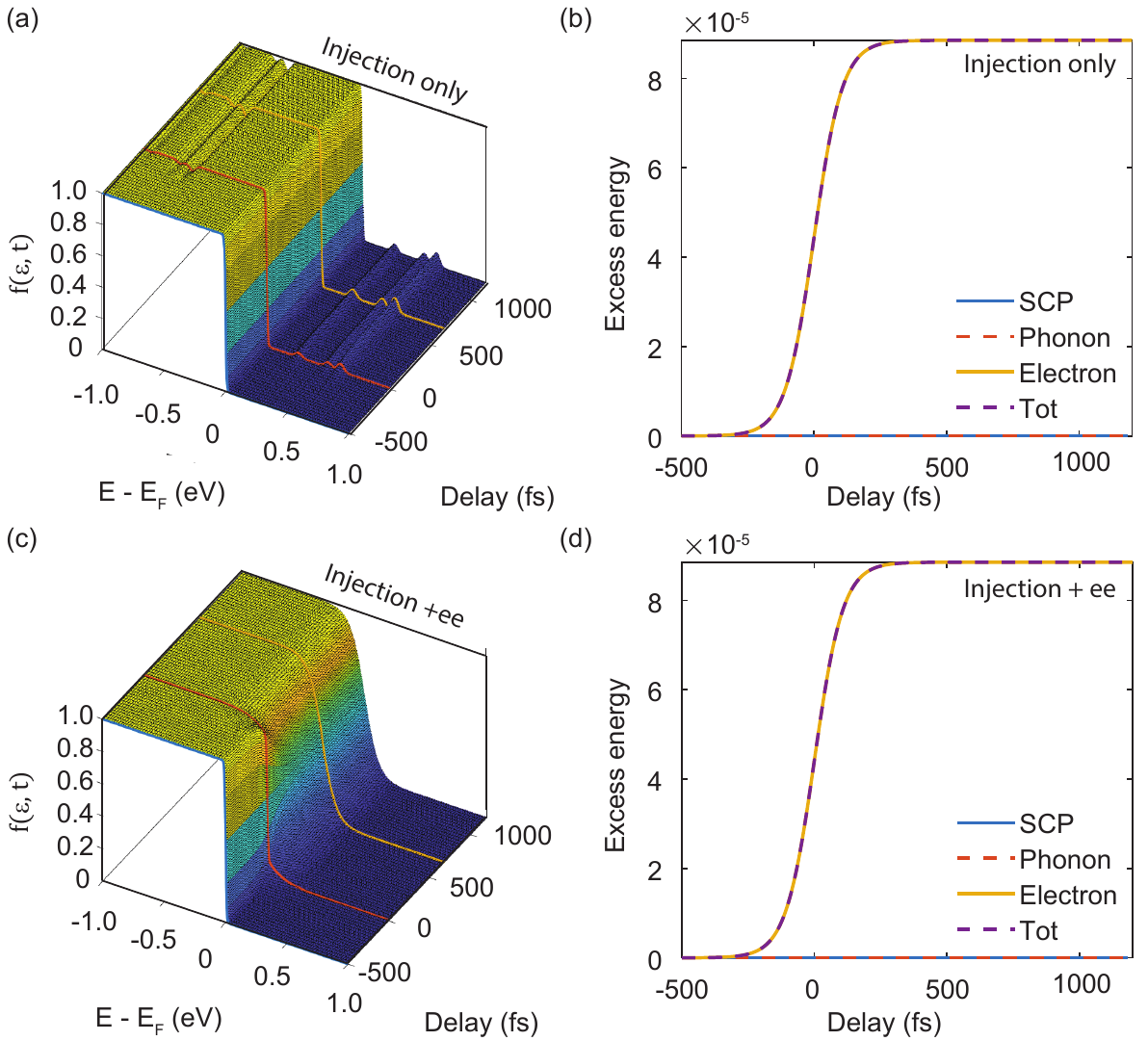}
    \caption{\textbf{Photo-excitation and e-e scattering} \textbf{(a), (c)} Occupation function simulated with only the injection term, and the injection plus e-e scattering, respectively. Delays at [-400, 0, 600]~fs are highlighted in blue, red, and yellow, respectively. \textbf{(b), (d)} The excess energy in the different subsystems as a function of delay. The solid yellow line gives the electron excess energy. The solid blue line gives the energy of the coupled phonon modes. The whole phonon subsystem is given by the dashed orange line. The dashed purple line gives the total energy of the system. There is no electron-phonon coupling; thus, excess energy in the phonon system is zero.}
    \label{SFig2}
\end{figure}
An example calculation using these two terms are shown in Fig.\,\ref{SFig2}. In Fig.\,\ref{SFig2}(a), we show the occupation function calculated with only the injection term; delays at -400~fs, 0~fs and 600~fs are highlighted. We see that the initial condition is just a Fermi-Dirac (FD) distribution at a temperature of $50~K$. At zero-delay (red), the population and depletion of states are visible as peaks. Since no scattering terms are included, the electrons stay in those states. The pump pulse is gone by approximately 600~fs (yellow), and the photo-excitation process is complete.

We check the energy of the system in Fig.\,\ref{SFig2}(b). The energy of each sub-system is defined as:
\begin{equation}
\begin{split}
    E_{\mathrm{e}}(t)&=\int d\epsilon  N(\epsilon)f(\epsilon,t)\epsilon;\\
    E_{\mathrm{ph}}(t)&=\int d\Omega  F(\Omega)n(\Omega,t)\Omega.\\
    E_{\mathrm{SCP}}(t) &= \int d\Omega'  F(\Omega')n(\Omega',t)\Omega',\quad\Omega'\subseteq \Omega \quad \mathrm{s.t.}\quad G(\Omega')\ne 0
\end{split}
\end{equation}
The excess energy is simply $E_{i}(t)-E_{i}(0)$. Since there is no electron-phonon (e-ph) coupling, there is no energy transferred to the phonon bath, so $E_{\mathrm{ph}}(t)=E_{\mathrm{SCP}}(t)$ is zero. Here we define $E_{\mathrm{SCP}}(t)$ as the excess energy of the phonon modes for which the e-ph coupling strength is non-zero. The excess energy in the electron bath is then equivalent to the total excess energy [see Fig.\,\ref{SFig2}(b)]. Next, we add the e-e scattering term. The occupation function is shown in Fig.\,\ref{SFig2}(c). As before, the initial condition is a FD at 50~K; however, at zero-delay, the injected carriers are no longer visible. We instead see that the electronic distribution has redistributed to form a FD at a higher temperature. The excess energy of the system is shown in Fig.\,\ref{SFig2}(d). Redistribution of the electron occupation does not change the energy between electron and phonon subsystems. Again, no energy is transferred to the phonon system, and the total excessive energy is equivalent to the electron excess energy.
\\
\textbf{Electron-phonon coupling:} The probability for an electron scattering with a phonon of energy $\Omega$ \textit{out} of a state $\epsilon$ is written as \cite{Sobota2014a}:
\begin{equation}
\begin{split}
\Gamma_{\mathrm{abs}}(\epsilon,\Omega)&=\frac{2\pi}{\hbar}|G(\epsilon, \Omega)|^2n(\Omega)F(\Omega)N(\epsilon+\Omega)(1-f(\epsilon+\Omega)),\\
\Gamma_{\mathrm{emi}}(\epsilon,\Omega)&=\frac{2\pi}{\hbar}|G(\epsilon, \Omega)|^2(1+n(\Omega))F(\Omega)N(\epsilon-\Omega)(1-f(\epsilon-\Omega)).
\end{split}
\end{equation}
Here $|G(\epsilon,\Omega)|^2$ is the coupling strength for an electron with energy $\epsilon$ with a phonon of energy $\Omega$, $n(\Omega)$ is the phonon occupation, $F(\Omega)$ is the phonon density of states and $f(\epsilon)$ is the electronic occupation.
Coupling to multiple phonons can be accounted for by an integral over $\Omega$:
\begin{equation}
\begin{split}
\left(\frac{\partial f^{\mathrm{out}}(\epsilon)}{\partial t}\right)_{\mathrm{e-ph}}=\int d\Omega \frac{F(\Omega)}{N(\epsilon)}[&\Gamma_{\mathrm{abs}}(\epsilon,\Omega)+\Gamma_{\mathrm{emi}}(\epsilon,\Omega)]N(\epsilon)f(\epsilon);\\
\left(\frac{\partial f^{\mathrm{in}}(\epsilon)}{\partial t}\right)_{\mathrm{e-ph}}=\int d\Omega \frac{F(\Omega)}{N(\epsilon)}[&\Gamma_{\mathrm{abs}}(\epsilon-\Omega,\Omega)N(\epsilon-\Omega)f(\epsilon-\Omega)\\
+&\Gamma_{\mathrm{emi}}(\epsilon,+\Omega,\Omega)N(\epsilon+\Omega)f(\epsilon+\Omega)].
\end{split}
\label{Eq: e_ph in}
\end{equation}
The phonon occupation $n(\Omega)$ is vanishingly small at an equilibrium temperature of 50~K, and initially, the $[1+n(\Omega)]$ emission term dominates. After optical excitation, $n(\Omega)$ increases substantially. We consider the time-evolution of $n(\Omega)$ in our simulation by rearranging Eq.\,\ref{Eq: e_ph in}, to obtain the phonon occupation rate equation:
\begin{equation}
\left(\frac{\partial n(\Omega)}{\partial t}\right)_{\mathrm{ph-e}}=\int d\epsilon \frac{N(\epsilon)}{F(\Omega)}[\Gamma_{\mathrm{emi}}(\epsilon,\Omega)-\Gamma_{\mathrm{abs}}(\epsilon,\Omega)]f(\epsilon).\\
\label{Eq: phonon occupation}
\end{equation}
\\
\textbf{Phonon-phonon scattering:} In the case that electrons couple strongly to few specific modes, these strongly-coupled modes increase quickly in occupation and subsequently decays into lower energy modes. The phonon-phonon (ph-ph) scattering term can be written as \cite{Ono2018}: 
\begin{equation}
    \begin{split}
        \left(\frac{\partial n(\Omega)}{\partial t}\right)_{\mathrm{ph-ph}}=\frac{2\pi}{\hbar}\Big[\frac{1}{2}&\int_{0}^{\Omega} d\xi |A(\Omega,\xi)|^2F(\xi)F(\Omega-\xi)\\
       \times[(1+n(\Omega))&n(\xi)n(\Omega-\xi)-n(\Omega)(1+n(\Omega-\xi))(1+n(\xi))]\\
        &+\int_{\Omega}^{\Omega_{\mathrm{max}}}d\xi |A(\Omega,\xi)|^2F(\xi)F(\xi-\Omega)\\ \times[(1+n(\Omega))&(1+n(\xi-\Omega))(n(\xi))-n(\Omega)n(\xi-\Omega)(1+n(\xi))]\Big]\\
    \end{split}
    \label{Eq: phph}
\end{equation}
\begin{figure}[t!]
    \centering
    \includegraphics{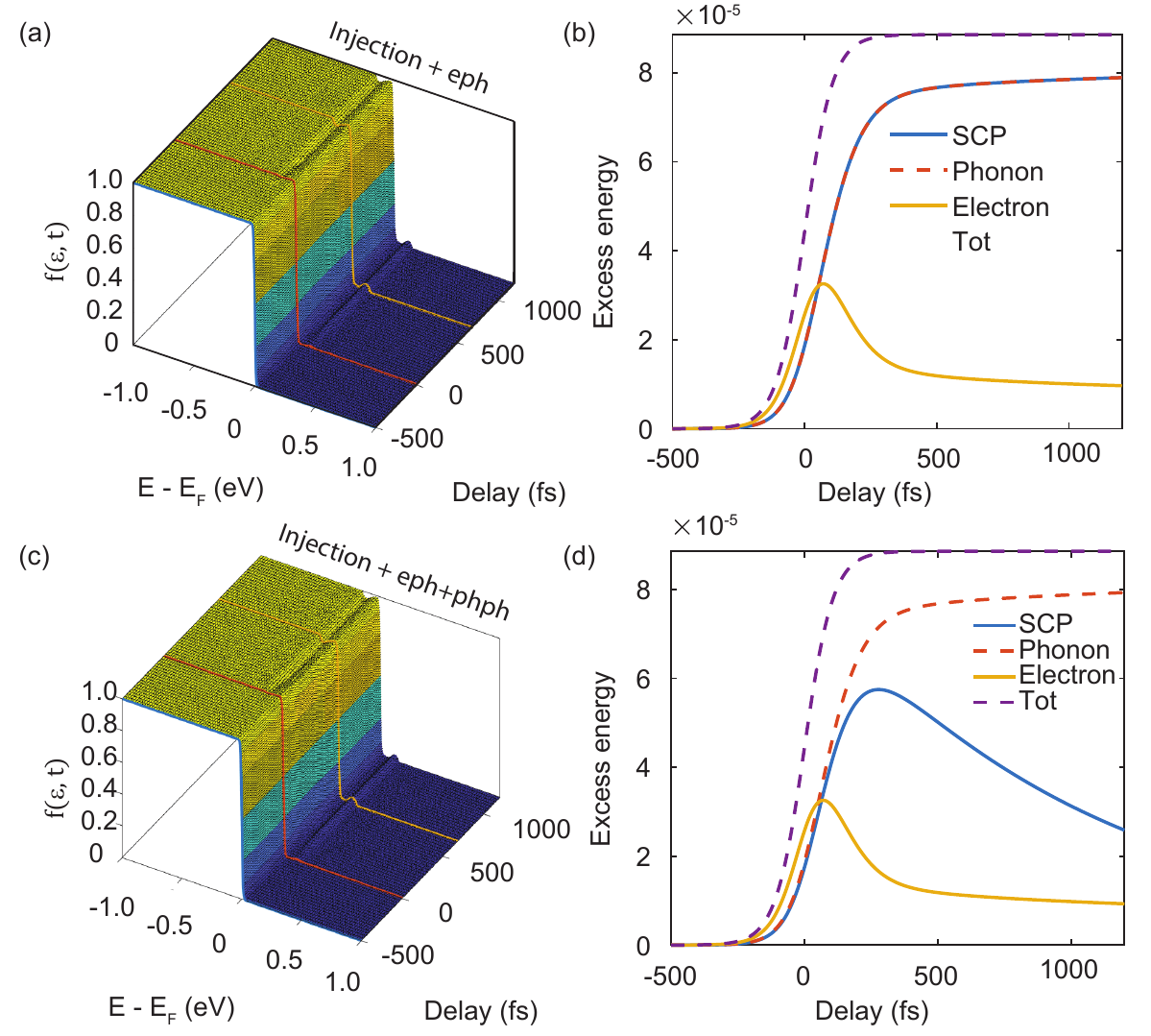}
    \caption{\textbf{Electron-phonon coupling and anharmonic phonon decay.} \textbf{(a), (c)} Simulation of $f(\epsilon, t)$ using the injection and electron-phonon (e-ph) terms and simulation using the injection, electron-phonon, and phonon-phonon (ph-ph) terms respectively. Delays at [-400, 0, 600]~fs are highlighted in blue, red, and yellow, respectively. \textbf{(b), (d)} The excess energy in the different subsystems as a function of delay. In (b), e-ph coupling allows the transfer of energy from the electron subsystem to the phonon subsystem, but the absence of ph-ph scattering means the excess energy of the strongly-coupled phonon (SCP) is equivalent to that of the entire phonon subsystem. In (d), turning on ph-ph scattering allows the SCP to decay into other phonon modes; thus excess energy of the SCP decreases, while the excess energy of the phonon bath remains equivalent to that in (b).}
    \label{SFig3}
\end{figure}
The combination of injection and e-ph scattering terms are shown in Fig.\,\ref{SFig3}(a). For simplicity, we show electrons coupling to just one phonon with energy $\Omega$, and label this SCP for ``strongly-coupled phonon". Similarly to Fig.\,\ref{SFig2}(c), the injected electrons rapidly relax from the high energy states towards the Fermi-level. However, rather than a FD distribution, we find  that electrons accumulate near $E_F$. This is a manifestation of bottle-necking of the relaxation channel due to Pauli-blocking, as discussed in the main text. Since we have not included e-e scattering or coupling to other phonon modes, this bottleneck prevents electrons from forming a FD distribution. We show the energy of the system in Fig.\,\ref{SFig3}(b), and observe that electrons quickly transfer energy to the phonon subsystem through the coupled mode. Since there is no anharmonic decay of the SCP, all the excess energy of the phonon bath is contained within the SCP, and the two quantities are equivalent. 

The ph-ph scattering term is added to the simulation along with e-ph scattering. The resultant $f(\epsilon, t)$ is given in Fig.\,\ref{SFig3}(c). The addition of this term does not affect the electronic distribution significantly, and so Fig.\,\ref{SFig3}(a) looks similar to Fig.\,\ref{SFig3}(b). The difference between the two simulations can be seen in the energy plot (Fig.\,\ref{SFig3}(d)). Now we see that energy is transferred from the electron subsystem to the coupled phonon subsystem. The energy then is transferred from this mode into the rest of the phonon subsystem. Thus, the orange dashed line is no longer equivalent to the solid blue line. We observe that the excess energy of the SCP decreases (blue line). The phonon bath,(which includes the SCP) remains the same as in panel (b), since the transfer of energy from electron bath to phonon bath through the SCP has not changed, and energy within the phonon bath is conserved.

\section{Boltzmann rate-equation model for graphite}
\label{App.Sim}
The equations used to generate Fig.\,2 in the main text are be written as:
\begin{subequations}
    \begin{alignat}{2}
        \begin{split}
        \Phi\left(\frac{\partial f}{\partial t}\right)_{\mathrm{inj}}=&\quad\left[\frac{1}{1.364\sigma_t}\mathrm{sech} \left(\frac{1.76x}{\sigma_t}\right)^2\right]\\
        &\quad\Bigg\{\frac{1}{\sqrt{2\pi}\sigma_{\epsilon}}\sum_i\Phi_i\left[\exp\left(-\frac{(\epsilon-\epsilon_{\mathrm{pop}}^i)^2}{2\sigma_\epsilon^2}\right)-\exp\left(-\frac{(\epsilon-\epsilon_{\mathrm{dep}}^i)^2}{2\sigma_\epsilon^2}\right)\right]\Bigg\}\label{Eq. Inj}
        \end{split}\\
        \begin{split}
        V\left(\frac{\partial f}{\partial t}\right)_{\mathrm{e-e}}=&\quad\frac{2\pi V}{\hbar}\int_0^{\eta_{\mathrm{max}}} d\eta\int d\epsilon'\left[N(\epsilon-\eta)N(\epsilon')N(\epsilon'+\eta)\right]\\
        &\quad\bigg\{f(\epsilon-\eta)f(\epsilon'+\eta)[1-f(\epsilon)][1-f(\epsilon')] -f(\epsilon)f(\epsilon')[1-f(\epsilon-\eta)][1-f(\epsilon'+\eta)]\bigg\}\label{Eq: ee}
        \end{split}\\
        \begin{split}
        G\left(\frac{\partial f}{\partial t}\right)_{\mathrm{e-ph}}=&\quad\frac{2\pi G }{\hbar}\int d\Omega F(\Omega)\bigg\{n(\Omega)\big([(1-f(\epsilon)]f(\epsilon-\Omega)N(\epsilon-\Omega)-f(\epsilon)[1-f(\epsilon+\Omega)]N(\epsilon+\Omega)\big)\\
        &\quad+[1+n(\Omega)]\big([(1-f(\epsilon)]f(\epsilon+\Omega)N(\epsilon+\Omega)-f(\epsilon)[1-f(\epsilon-\Omega)]N(\epsilon-\Omega)]\big)\bigg\}\label{Eq: eph}
        \end{split}\\
        \begin{split}
        G\left(\frac{\partial n}{\partial t}\right)_{\mathrm{ph-e}}=&\quad\frac{2\pi G}{\hbar}\int d\epsilon N(\epsilon)f(\epsilon)\bigg\{[1+n(\Omega)][1-f(\epsilon-\Omega)]N(\epsilon-\Omega)-n(\Omega)[1-f(\epsilon+\Omega)]N(\epsilon+\Omega)\bigg\}\label{Eq: phe}
        \end{split}\\
        \begin{split}
        A\left(\frac{\partial n}{\partial t}\right)_{\mathrm{ph-ph}}=&\quad\frac{2\pi A}{\hbar}\Bigg[\frac{1}{2}\int_{0}^{\Omega} d\xi F(\xi)F(\Omega-\xi)\quad\bigg\{[1+n(\Omega)n(\xi)n(\Omega-\xi)-n(\Omega)[1+n(\Omega-\xi)[1+n(\xi)]]\bigg\}\\
        &\quad\int_{\Omega}^{\Omega_{\mathrm{max}}}d\xi F(\xi)F(\xi-\Omega)\bigg\{[1+n(\Omega)][1+n(\xi-\Omega)](n(\xi))-n(\Omega)n(\xi-\Omega)[1+n(\xi)]\bigg\}\Bigg]\label{Eq: phph2}
        \end{split}
    \end{alignat}

\end{subequations}
We adopt a quasi-phenomenological approach to the Boltzmann rate-equation model. As we want to explore the effect of occupation function on the evolution of TR-ARPES spectra, we set the e-e, e-ph, and ph-ph scattering potentials to a constant; $V$, $G$, and $A$ respectively. In the model presented in Fig.\,2 of the main text, the values used were: $G^2=0.07$, $V^2=1.2\times10^4$, $A=1\times10^{-4}$, and $\Phi=1\times 10^{-3}$.

The injection term is given by a sech function in time and energy, with pulse duration $\sigma_t$ and bandwidth $\sigma_\epsilon$. Carriers are injected at states $\epsilon_{\mathrm{pop}^i}$, and depleted at states $\epsilon_{\mathrm{dep}^i}$. Following the observed transitions in a previous study, we populate (deplete) electrons into states $[0.26, 0.5, 0.6]$~eV ($[-0.94, -0.7, -0.6])$~eV \cite{Na2019}. $\Phi_i$ controls number of excited electrons, or the amount of spectral weight transfer. In Eq.\ref{Eq: ee}, the integral over $\eta$ accounts for all possible energy transfers in e-e scattering events, and the integral over $\epsilon'$ accounts for elastic scattering, such that energy within the electron bath is conserved. The e-ph and ph-e terms allow for energy transfer from the electron to the phonon bath, and conserves the energy of the entire system. The ph-ph term then redistributes energy within the phonon bath, such that it evolves towards a Bose-Einstein distribution. The electron density of states have been computed from a tight-binding model from a previous work \cite{Na2019}. The phonon density of states is taken from Ref. \cite{Tohei2006}. although the coarse grid over phonon energies makes the model fairly insensitive to the phonon density of states.

The strongest optical transition energies for a $1.2$~eV pump pulse is given by an optical-joint density of states calculation in a previous work \cite{Na2019}. Characteristic pulse bandwidth ($\sigma_{\epsilon}$) and duration ($\sigma_t$) are informed by experimental parameters ($190$~fs, $30$~meV). We limit the e-ph scattering term to the $\mathrm{A_1'}$ SCP, which anharmonically decays into low energy phonon modes.

\begin{figure}[t!]
    \centering
    \includegraphics{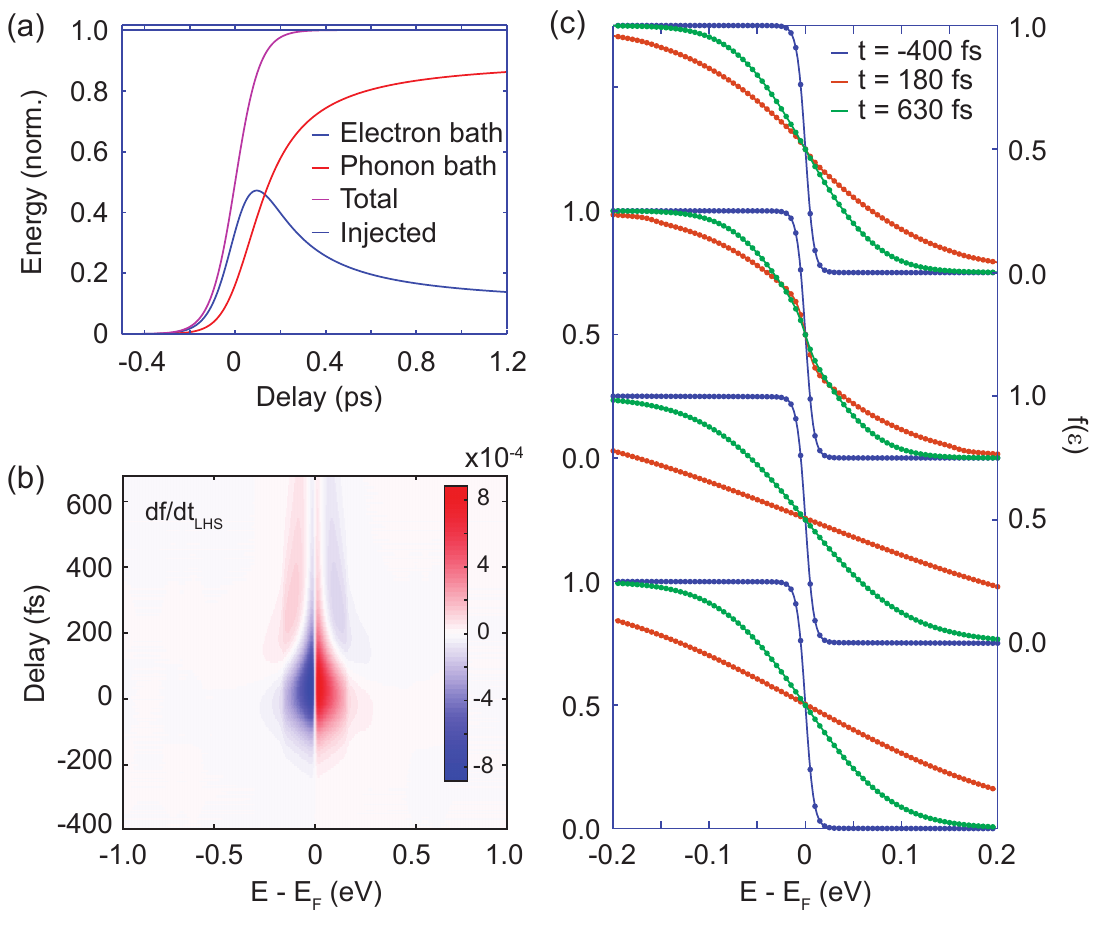}
    \caption{\textbf{Simulation checks} (a) The energy of the electron and phonon subsystems in the simulation as a function of pump-probe delay are shown in blue and red solid lines. The system's total energy converges to that injected by the pump pulse (with no scattering processes). (b) The left-hand-side (LHS) of the Boltzmann rate equations (Eq.\,3). Carriers are injected/depleted near ~0.6 eV. The dynamics are well contained within the energy range $[-1,1]$~eV. (c): The occupation function at -400~fs, 180~fs, and 630~fs, calculated with a mesh of 5~meV (markers) and 2~meV (line). The two overlap, showing a good convergence.}
    \label{SFig4}
\end{figure}
\section{Simulation checks}
\label{App.Checks}
To ensure the accuracy and correctness of the simulation, we check that:
\begin{itemize}
    \item energy is conserved within the system
    \item the energy domain considered is large enough to include all relevant scattering processes
    \item the energy grid is dense enough to reach convergence
\end{itemize}

\textbf{Energy conservation:} We first simulate the occupation function, including \textit{only,} the injection term. The energy $\Delta E=E(t)-E(-\infty)$ injected into the electron system is the solid black line in Fig.\,\ref{SFig4}(a). Next, we turn on scattering processes, keeping the same pump fluence. 

The total excess energy $\Delta E(t)$ is then given by $\Delta E_\mathrm{e}+\Delta E_\mathrm{ph}$. $\Delta E_{\mathrm{e}}$ and $\Delta E_{\mathrm{ph}}$ are red and blue solid lines in Fig.\,\ref{SFig4}(a). The total energy $\Delta E$ is the solid magenta line. This value is directly compared with the injected energy (black line). As we can see, the total energy of the simulation converges to the injected energy, so our scattering terms maintain energy conservation. 

\textbf{Scattering rate:} To ensure we include all relevant scattering processes, we look directly at the scattering rate. The energy range should encompass all states affected by the OJDOS, but electrons can scatter into states above/below those states. The balance is to include all necessary states without incurring unnecessary computation cost. The left-hand-side (LHS) of Eq.\,3, $\partial f(\epsilon,t)/\partial t$ is shown in Fig.\,\ref{SFig4}(b). By visual inspection, the terms contributing to $\partial f/\partial t$ are contained in the energy domain $[-1,1]$~eV. Note: as fluence increases, the electron distribution spreads over a larger energy domain. Thus, the energy domain needs also increase to conserve the energy of the system and include the relevant scattering processes.

Lastly, we check the convergence of the simulation. In Fig.\,\ref{SFig4}(c), we compare $f(\epsilon,t)$ at t=[-400, 180, 630]~fs for a simulation with 5~meV mesh size (circles) and 2~meV mesh size (line). We see the two overlap quite well; the error between these two simulations is less that one percent of $f(\epsilon, t)$.

\begin{figure}
    \centering
    \includegraphics{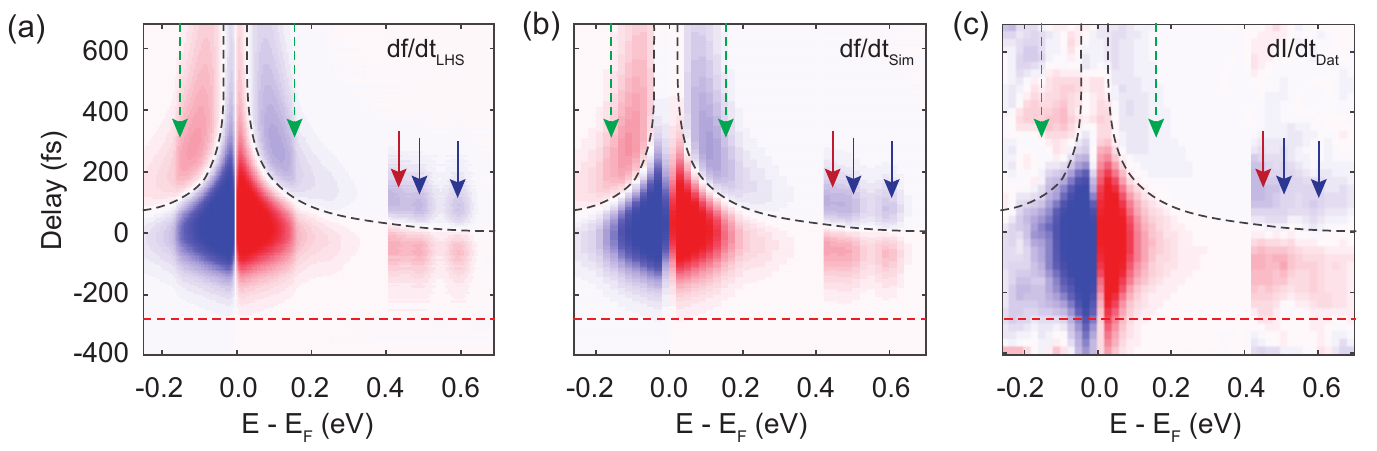}
    \caption{\textbf{Electron scattering rate calculated from data and experiment} (a) $\partial f/\partial t$ calculated from Eq.\,3 in the main text. (b) $\partial f/\partial t$ is calculated from the simulation. The output of the simulation $f(\epsilon,t)$ is integrated into bins 50~meV wide, the tangent for each delay is extracted from a moving fit of the intensity in a 50~fs window with a 1st order polynomial. (c) $dI/dt$ calculated from the data. The intensity is integrated into bins 50~meV wide, and the tangent for each delay is extracted from a moving fit of the intensity in a 50~fs window with a 1st order polynomial.  The dynamics are contained within the energy range $[-1,1]$~eV. Arrows indicate non-thermal features, including the direct transition at 0.6~eV and 0.5 eV (blue) and the phonon-induced replica at 0.44 eV (red). Green dashed arrows mark the energy domain $\pm \hbar\Omega$. In all panels, the region between 0.4~eV and 0.7~eV is enhanced by a factor of 20 for visualization purposes.}
    \label{SFig5}
\end{figure}

\section{Simulation and Analysis comparison}
\label{App.Comp}
To match the simulation to the data, we vary four parameters: The fluence $\Phi$, and scattering strengths $G$, $V$, and $A$. We also define the pulse shape, pulse bandwidth (informed by the experiment), OJDOS (informed by experiment and calculations), the electron and phonon DOS (given by calculations). 

To avoid over parameterization, we set $G$, $V$, and $A$ to constants. The DOS is calculated from a tight-binding model consistent with previous work, and the OJDOS is given by experimental observations in that same data set\cite{Na2019}. The time-domain pulse shape is simply a sech function with an FWHM equal to the measured system time resolution.

The data set shown in the main text spans the delays $[-400, 680]$~fs. In this time domain, the electronic distribution is determined by the interplay between e-e scattering $V$, e-ph scattering $G$, and the fluence $\Phi$. The phonon density of states and the anharmonic scattering strength $A$ affect the distribution on much longer timescales and does not play a prominent role here. We compare high-fluence simulations with high-fluence experiments measured over the picosecond timescale to qualitatively determine $A$ \cite{Johannsen2013, Gierz2013c, Ulstrup2015, Stange2015}. In this next section, we discuss comparisons with data used to pin the parameters of the simulation that were not shown in the main text.

Firstly, we directly compare the total rate-of-change of the occupation function $\partial f\partial t$ between simulation and data. In the experimental data, the rate-of-change of the ARPES intensity $dI/dt$ takes the place of $\partial f/\partial t$, given the assumptions discussed in the main text (Eq.\,2). We can obtain this quantity by integrating $I$ in a small energy domain, such that $I(t)$ is reasonably smooth, then taking the tangent of the curve at each delay. In Fig.\,\ref{SFig5}(a) and (b), we shows $(\partial f/\partial t)$ calculated directly from Eq.\,3 of the main text and $(\partial f/\partial t)$ produced by calculating the tangent of $f(\epsilon,t)$ at each delay. We see the two are exactly equal, except that the resolution in panel (b) is reduced as a result of integration over a finite energy domain.  

We now apply this analysis to the data, shown in Fig.\,\ref{SFig5}(c). For the majority of delays, we observe a good agreement between the data and the simulation, including:
\begin{itemize}
    \item the observation of the direct-transitions at 0.6~eV and 0.5~eV (blue arrow)
    \item the phonon-induced replica at 0.44~eV (red arrow)
    \item the accumulation of electrons outside the phonon window (dashed green arrows)
    \item the sign reversal both above and below the Fermi-energy (black dashed lines)
\end{itemize}  
The main difference lies in the dynamics before -300~fs, indicated by dashed red lines. We assumed a sech pulse shape with an FWHM of 190~fs, determined from high-statistics measurements of the intensity of the direct-transition peak at 0.6~eV \cite{Mills2018}. For this pulse duration, we do not expect significant pump excitation (and by extension, electron dynamics) before -300~fs. Dynamics before -300~fs can only be explained by a low amplitude tail in the pump pulse. In the following figures, the effects of this tail consistently show up. However, as we are primarily interested in the physics post-pump excitation, we do not expend effort in modelling this tail.
\begin{figure}
    \centering
    \includegraphics{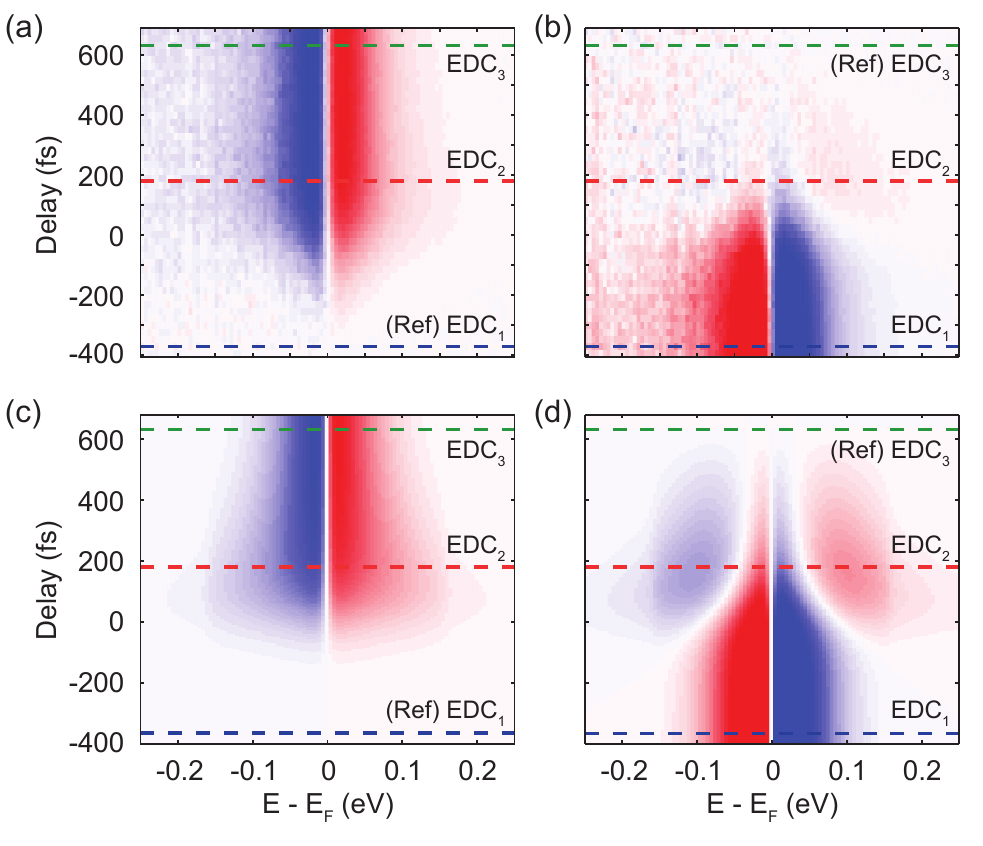}
    \caption{\textbf{Differential maps} (a, b) $\Delta\int_k\mathrm{EDC}$ maps computed using the Ref$_1$ (Ref$_3$), at delays -400~fs (630~fs) respectively. Blue (red) represent negative (positive) values. (c, d) $\Delta f$ maps using the output of the simulation in Fig.\,2 in the main text. Panel (c) and (d) are computed using $f(\epsilon, -400~\mathrm{fs})$ and $f(\epsilon, 630~\mathrm{fs})$, as reference.}
    \label{SFig6}
\end{figure}

In the main text, we showed an unconventional differential momentum-integrated energy distribution curves ($\Delta\int_k\mathrm{EDC}$)s, which use the delay at t=630~fs as a reference, rather than the typical unpumped distribution as a reference. Here we show the differential maps at all delays, comparing side by side the use of both references in the ARPES data and the simulated $f(\epsilon, t)$.  $\Delta\int_k \mathrm{EDC}$ maps calculated from the ARPES ata using Ref$_1$ (-400~fs) and Ref$_2$ (630~fs) are shown in Fig.\,\ref{SFig6}(a) and (b) respectively. In panel (a), we see the depletion (population) below (above) the Fermi energy that is commonly associated with thermal broadening. The single sign change that arises from by taking the difference of two FD distributions at different temperatures manifests here as the white colour transition between blue and red. In panel (b), the multiple sign changes that signify a non-thermal distribution manifests as multiple colour changes. At 180~fs (indicated by the red dashed line), starting from negative energies, we see the colour change from blue to red, to blue, and back to red again. Simulated $\Delta f(\epsilon,t)$ maps are similarly shown in Fig.\,\ref{SFig6}(c) and (d). The $\Delta f$ map using Ref$_1$ highlights electron dynamics at early delays. The discrepancy here is again due to the tail of the pump pulse, as we saw earlier. The $\Delta f$ map using Ref$_3$ instead highlights non-thermal electron dynamics near zero-delay. In this region, we see a good agreement between the data and the simulation.
\begin{figure}
    \centering
    \includegraphics{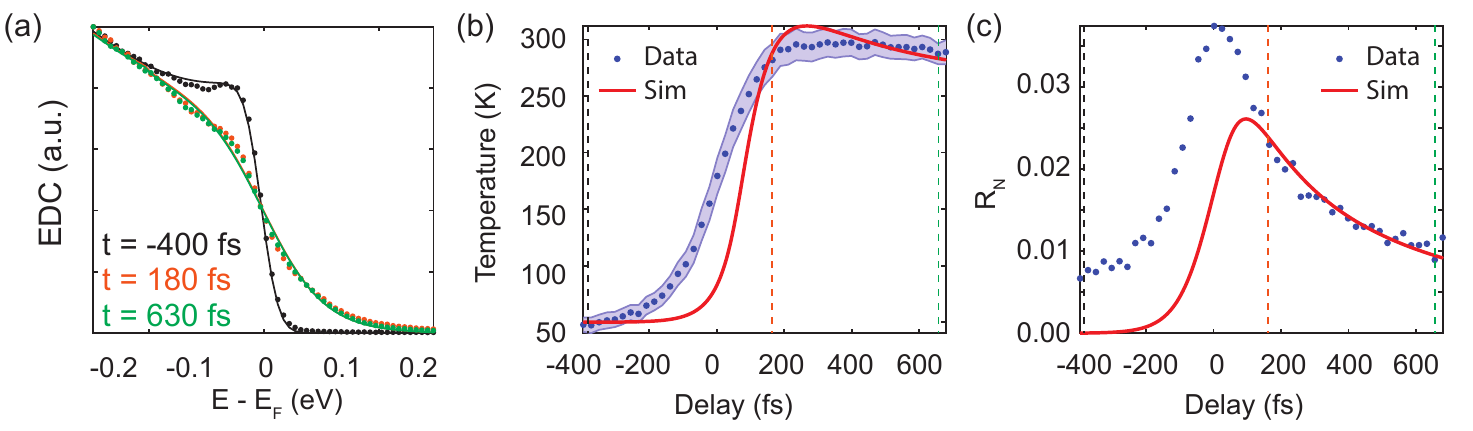}
    \caption{\textbf{Temperature fits of the data and simulation} (a) EDCs at delays [-400, 180, 630]~fs (markers) and the phenomenological fit model (lines). (b) Temperature extracted from the fit of the data in panel (a) (blue markers) and from Fermi-Dirac fits of the simulation (red line). Delays shown in panel (a) are indicated by dashed lines in the corresponding color. (c) The normalized residuals ($R_N$) from the effective temperature fit of data (blue markers) and the simulation (red line).}
    \label{SFig7}
\end{figure}   

Lastly, we discuss the temperature fit of the data. The fit model and the data are shown in Fig.\,\ref{SFig7} at three different delays. As discussed in the main text, we cannot determine whether the residuals of the fit are due to model imperfection. The effective temperature and normalized residuals $R_N$ are shown in the blue markers of Fig.\,\ref{SFig7}(b), and (c), respectively. The temperature fit of the simulation by using a FD is shown in red lines. We see that in comparison to the simulation, the electronic temperature increases much faster at negative delays, and has a higher $R_N$ value. Both of these are signatures of the tail of the pump pulse, which is not present in the simulation. Past 180~fs, both quantities are well reproduced. Therefore, although $R_N$ obtained from the data captures both model imperfections as well as non-thermal features, it can be used as a diagnostic for whether electronic distributions are thermal.
\twocolumngrid

\end{document}